# AI Delivers Creative Output but Struggles with Thinking Processes

Man Zhang [a, b,1], Ying Li [a, b,1], Yang Peng [a, b], Yijia Sun [a, b], Wenxin Guo [a, b], Huiqing Hu [a, b,*], Shi Chen [c, d,*], Qingbai Zhao [a, b,*]

[a] Key Laboratory of Adolescent Cyberpsychology and Behavior (CCNU), Ministry of Education, Wuhan 430079, China;

[b] Key Laboratory of Human Development and Mental Health of Hubei Province, School of Psychology, Central China Normal University, Wuhan 430079, China;

[c] School of Medical Humanities, Hubei University of Chinese Medicine, Wuhan 430065, China；

[d] Hubei Shizhen Laboratory, Wuhan 430000, China.

[1] These authors contributed equally to this work and should be considered co-first authors.

[*] **Corresponding author:**

Email: zqbznr@ccnu.edu.cn (Qingbai Zhao); chs22@hbucm.edu.cn (Shi Chen);

huiqinghu@ccnu.edu.cn (Huiqing Hu).

# Abstract

A key objective in artificial intelligence (AI) development is to create systems that match or surpass human creativity. Although current AI models perform well across diverse creative tasks, it remains unclear whether these achievements reflect genuine creative thinking. This study examined whether AI models (GPT-3.5-turbo, GPT-4, and GPT-4o) engage in creative thinking by comparing their performance with humans across various creative tasks and core cognitive processes. Results showed that AI models outperformed humans in divergent thinking, convergent thinking, and insight problem-solving, but underperformed in creative writing. Compared to humans, AI generated lower forward flow values in both free and chain association tasks and showed lower accuracy in the representational change task. In creative evaluation, AI exhibited no significant correlation between the weights of novelty and appropriateness when predicting creative ratings, suggesting the absence of a human-like trade-off strategy. AI also had higher decision error scores in creative selection, suggesting difficulty identifying the most creative ideas. These findings suggest that while AI can mimic human creativity, its strong performance in creative tasks is likely driven by non-creative mechanisms rather than genuine creative thinking.

Keywords: Creative thinking；  Artificial intelligence；  Human–AI comparison.

# 1 Introduction

Creative thinking enables humans to transcend the boundaries of knowledge and experience, allowing them to imagine and create things previously unseen. It supports a wide range of achievements, from devising new recipes to inventing groundbreaking technologies. Creative thinking is essential for both individual success and societal progress. Artificial Intelligence (AI) has demonstrated its capability to solve creative problems (Raz et al., 2024; Hubert et al., 2024; Boussioux et al., 2024) and even generate more creative research ideas than human researchers (Si et al., 2024). The continuous development of AI models has drawn increasing attention to whether AI has creative thinking.

AI models, such as ChatGPT (Generative Pre-trained Transformer), are trained on large-scale language datasets using methods like "reinforcement learning from human feedback" (Vaswani et al., 2017; Ouyang et al., 2022), which enable these models to generate creative ideas (Runco, 2023). However, it remains unknown whether the creative ideas generated by AI models truly arise from genuine creative thinking or from some non-creative mechanisms, such as searching and integrating existing ideas (Zhao et al., 2024; Ismayilzada et al., 2024). Clarifying this ambiguity requires examining not only performance outcomes but also the thinking processes underlying these creative outputs (Runco, 2023; Aru, 2025; Franceschelli & Musolesi, 2024a). Therefore, this study investigates two critical questions: First, on which creative tasks can AI outperform humans? Second, can AI outperform humans in the core thinking processes that underlie these creative tasks? To address these questions, we conducted eight experiments systematically comparing human and AI performance across diverse creative tasks and essential creative thinking processes.

## 1.1 Evaluating the creative potential of AI

Numerous studies have investigated the differences between AI and humans in performing creative tasks, mainly encompassing divergent thinking, convergent thinking, insight problem-solving, and creative writing. Initially, research primarily used divergent thinking tasks, which require participants to generate various creative ideas (Stevenson et al., 2022; Guzik et al., 2023; Koivisto & Grassini, 2023; Cropley, 2023). While early findings indicated that the GPT-3 model did not surpass human performance (Stevenson et al., 2022), subsequent evaluations showed that GPT-4 excelled in various tasks such as the alternative uses task (AUT) and divergent association task (Koivisto & Grassini, 2023; Hubert et al., 2024; Cropley, 2023; Bellemare-Pepin et al., 2024). However, these studies also found significant homogeneity in creative performance across these AI models (Hubert et al., 2024; Zhou et al., 2024; Wenger & Kenett, 2025). For example, AI exhibited a significantly higher response repetition rate than humans in the divergent association task (Niloy et al., 2024; Hubert et al., 2024). This lack of variation contradicts the fundamental aim of creative thinking, which values diversity and novelty (Johnson et al., 2021).

Recent studies have explored AI performance in convergent thinking tasks requiring identifying novel and meaningful relationships between concepts (Abraham, 2018). Previous research found that AI models such as GPT-3.5 and LLaMa performed significantly worse than humans in assessing conceptual similarities in a story analogy reasoning task (Jiayang et al., 2023). Another study tested five AI models, including GPT-4Omni and Claude 3.5 Sonnet, on a word classification game that required convergent thinking. The results showed that while AI performed well in semantic categorization, it struggled with unconventional classifications requiring encyclopedic knowledge (Samadarshi et al., 2024). Similarly, Tian et al. (2023) introduced the MACGYVER problem set, which evaluates divergent and convergent thinking through real-world problem-solving tasks. Their findings showed that AI models frequently generated unrealistic or incorrect solutions, and failed to apply knowledge flexibly. These findings suggested the limitations of AI models in identifying novel and meaningful relationships between concepts.

Insight problems may also present challenges for AI models. These problems typically require sudden cognitive restructuring rather than gradual reasoning or trial-and-error approaches (Duncker, 1945; Ellen, 1982; Ohlsson, 1992; Knoblich et al., 1999). The key to solving such problems requires abandoning an initially incorrect representation and discovering a new representation that reconstructs the elements in the problem (Wiley & Danek, 2023). For example, when solving a riddle, insight occurs when transitioning from a literal interpretation to an abstract, metaphorical, or symbolic understanding to reach a solution. While AI models excel in solving complex tasks through probabilistic prediction and contextual reasoning (Hagendorff et al., 2023; Si et al., 2024), they may struggle with insight problems. However, empirical findings suggest that AI models outperform humans in solving insight problems. Orrù et al. (2023) reported that GPT-3 performed comparably to humans on verbal insight tasks, while Raz et al. (2024) found that GPT-4 significantly outperformed humans on similar tasks. These findings raise further questions about whether AI's high accuracy in insight problem-solving stems from specialized algorithms and vast datasets or it reflects the creative thinking required for such tasks.

Moreover, research indicates that AI performs relatively poorly in creative writing tasks (Orwig et al., 2024; Sun et al., 2024; Bellemare-Pepin et al., 2024). For example, Orwig et al. (2024) found that stories generated by GPT-4 were significantly less creative than those produced by humans. When comparing GPT-4 with expert writers, researchers observed that GPT-4's creative performance was significantly worse than that of professionals (Marco et al., 2024). Similarly, Gómez-Rodríguez and Williams (2023) reported that GPT models scored significantly lower than professional writers in a novel writing task. Furthermore, studies also reported that AI-generated texts exhibit high content homogeneity (Padmakumar & He, 2024; Sun et al., 2024).

Taken together, AI surpasses humans in some creative tasks but struggles in others, raising questions about whether it truly engages in creative thinking. Researchers increasingly argue that AI's performance on these tasks alone is insufficient to settle this debate (Aru, 2024; Franceschelli & Musolesi, 2024a; Wingström et al., 2022). To gain a deeper understanding, it is essential to examine the underlying processes that drive these creative outcomes (Runco, 2023).

## 1.2 The present study

This work examines differences between AI and humans from two distinct perspectives: performance on various classic creative tasks, and performance in core creative thinking processes. We selected three representative AI models (GPT-3.5, GPT-4, and GPT-4o) due to their widespread applicability and cost-effectiveness (Reinhart et al., 2025). GPT-4 and GPT-4o have also demonstrated high and consistent performance across various AI benchmark tests (Chiang et al., 2024).

Study 1 systematically evaluated performance differences between AI and humans across four classic creative tasks. Specifically, Experiment 1 used a divergent thinking task (Alternate Uses Task); Experiment 2 used a convergent thinking task (Remote Associates Test); Experiment 3 assessed insight problem-solving through two Chinese verbal tasks (Chinese character puzzles, the stumper task), and an English task that combines visual and verbal cues (Rebus); and Experiment 4 comprised two creative writing tasks (the five-sentence story task and the advertising creative writing task).

Study 2 further examined the difference between AI models and humans across four essential creative thinking processes that support creative outputs: association (Mednick, 1962; Beaty & Kenett, 2023), representational change (Ohlsson, 1992), creative evaluation (Ellamil et al., 2012), and creative selection (Rietzschel et al., 2024). Specifically, Experiment 5 employed a free association task and chain association to measure how humans/AI models extracted information to avoid repetitions or common responses; Experiment 6 used a representational change task to evaluate whether humans/AI models engaged in cognitive restructuring of the problem representation; Experiment 7 focused on how humans/AI models would weigh novelty and appropriateness when evaluating idea's creativity in a creative evaluation task. Experiment 8 examined their ability to select the most creative idea from multiple ideas.

# 2 study 1

## 2.1 Experiment 1

### 2.1.1 Participant

In Experiment 1, 100 Chinese participants were recruited (Mage = 20.52 years; 51.00% female). AI-generated data were obtained from three ChatGPT versions: GPT-3.5, GPT-4o, and GPT-4. For each GPT model, 100 instances' data were collected, with each instance's data from a single task run following the task instructions and parameter settings. Thus, 300 GPT instances' data were collected across the three models. The parameter settings followed prior studies (Orwig et al., 2024; Stevenson et al., 2022) and were applied in all experiments: temperature (ranging from 0.60 to 0.80) for output randomness, frequency penalty (set to 1), and presence penalty (set to 1). This study was approved by the university's Institutional Ethics Review Board and was preregistered on the Open Science Framework (OSF).

### 2.1.2 Task and measures

***Alternative Uses Task (AUT).*** Experiment 1 used the classic Alternative Uses Task (AUT) to assess divergent thinking performance. Human participants were presented with two common objects ('newspaper' and 'brick') and instructed to generate as many creative uses as possible within two minutes for each item. To ensure comparability, the number of ideas generated by each human participant was equally matched (1:1) with responses from the three GPT models (Hubert et al., 2024). For instance, if human participant #1 generated five ideas, the corresponding GPT-3.5, GPT-4, and GPT-4o instances (#1) were instructed to generate five responses.

***Measures.*** Responses were scored for novelty, appropriateness, and flexibility. Six raters used a categorical 7-point scale to rate novelty (whether the response was unique, uncommon, and rarely thought of, from low to high novelty), appropriateness (whether the response was valuable, meaningful, and feasible, from low to high appropriateness), and flexibility (number of idea categories). Each object was rated by three raters: brick (novelty: $\alpha = 0.81$, appropriateness: $\alpha = 0.82$, flexibility: $\alpha = 0.96$) and newspaper (novelty: $\alpha = 0.95$, appropriateness: $\alpha = 0.89$, flexibility: $\alpha = 0.98$). We calculated each participant/instance's mean score across the two objects.

### 2.1.3 Procedure

***Human procedure.*** Data from human participants were collected through offline and online experiments. In the offline experiments, participants read the instructions, signed an informed consent form at the lab, and began the Alternative Uses Task (AUT). The formal experiment involved two common objects (*'newspaper'* and *'brick'*), with one object randomly presented per trial. Participants were instructed to generate as many creative uses as possible and enter their responses via the keyboard.

For the online experiment, participants reviewed the electronic informed consent form before starting and explicitly consented to audio and video recordings throughout the experiment. To enhance the procedural rigor of processes, participants were asked to join a Tencent meeting with their camera and microphone turned on. The camera was positioned to capture the computer screen and keyboard. The task content, stimulus materials, trial order, and time limits in the online experiment were identical to those in the offline experiment.

***AI procedure.*** After human data collection, we used a Python script to call the OpenAI API, gathering responses from GPT-3.5-turbo, GPT-4o, and GPT-4 for the AUT. Each AI model generated responses based on a system prompt and a user prompt. System prompts closely aligned with human task instructions, incorporating minor additions to control response quantity and prevent irrelevant outputs. To ensure comparability, the user prompt explicitly requested the same number of responses as generated by corresponding human participants (e.g., "Please provide {number} creative uses for {item}," where "{number}" corresponds to the human response count, and "{item}" is the stimulus object). All detailed human instructions and AI prompts are provided in Supplementary Information

B.

### 2.1.4 Results

The results showed that idea novelty in humans was significantly lower than in the three GPT models. Specifically, GPT-3.5-turbo ($t$= 5.32, 95% CI = [-0.54, -0.19], $p$ < 0.001), GPT-4 ($t$= 5.10, 95% CI = [-0.51, -0.16], $p$ < 0.001), and GPT-4o ($t$= 8.89, 95% CI = [-0.76, -0.42], , $p$ < 0.001) all got significantly higher novelty scores than humans. In contrast, the appropriateness scores for human-generated ideas were significantly higher than those for GPT-3.5-turbo ($t$=-3.37, 95% CI = [0.05, 0.37], $p$ = 0.005) and GPT-4 ($t$=-4.60, 95% CI = [0.12, 0.43], $p$ < 0.001). However, no significant difference in the appropriateness scores was found between humans and GPT-4o ($p$ = 0.163). Flexibility scores for humans were significantly lower than those of all GPT models, including GPT-3.5-turbo ($t$=-7.86, 95% CI = [-0.26, -0.13], $p$ < 0.001), GPT-4 ($t$=-6.49, 95% CI = [-0.23, -0.09], $p$ < 0.001), and GPT-4o ($t$=-4.41, 95% CI = [-0.17, -0.04], $p$ < 0.001). Figure 1 illustrates the differences between humans and the three GPT models across these metrics. Descriptive statistics for all metrics are shown in Table S1.

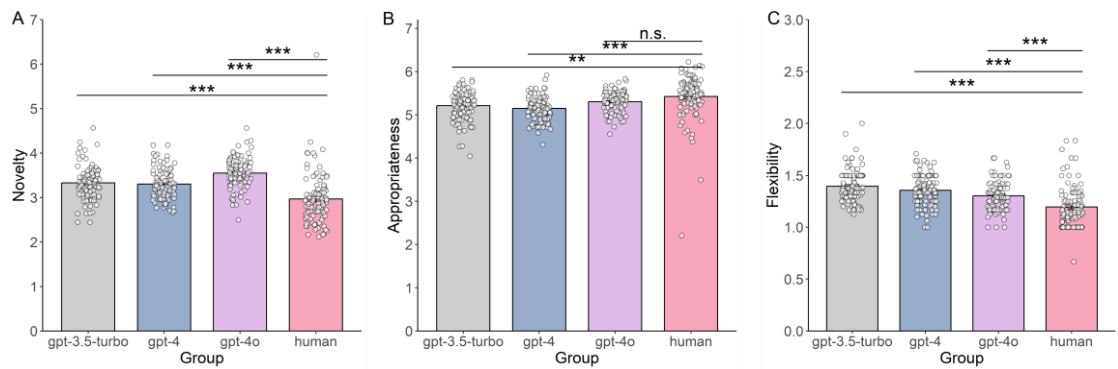

**Figure 1.** Comparison of novelty ratings (A), appropriateness (B), and flexibility (C) across GPT-3.5-turbo, GPT-4, GPT-4o, and human groups. Note: **$p$ < .01, ***$p$ < .001.

## 2.2 Experiment 2

### 2.2.1 Participant

We recruited 61 human participants (*Mage* = 20.52 years; 54.10% female) to complete the Remote Associates Test (RAT). For each GPT model (GPT-3.5-turbo, GPT-4o, and GPT-4), we collected 61 instances' data.

### 2.2.2 Task and measures

*Remote Associates Test (RAT).* The RAT is a widely used measure of convergent thinking

(Mednick, 1962). In each RAT trial, participants were presented with three words and had to generate a fourth word that connects or fits all three within one minute (e.g., "cream, baking, wedding," with the correct answer being "cake"). The task consists of 30 questions, evenly distributed across three levels of difficulty: easy, medium, and hard.

*Measures*. The task performance was assessed by accuracy. Additionally, we quantified the relative semantic similarity (RSS) to analyze patterns in incorrect responses. RSS was calculated by dividing the average semantic similarity between each incorrect response and the three stimulus words by the average semantic similarity between the correct answer and these words. Semantic similarity scores were computed using data from the Tencent corpus, which comprises over 12 million Chinese words.

### 2.2.3 Procedure

*Human procedure.* Experiment 2 used offline and online data collection and followed the same pre-experiment procedures as Experiment 1. Participants completed the RAT, in which they were presented with 30 sets of three words, each displayed in random order. For each set, participants had one minute to enter a single word connecting all three presented words.

*AI procedure*. AI data collection (GPT-3.5-turbo, GPT-4o, and GPT-4) followed Experiment 1's prompt structure. The system prompts closely matched the human task instructions, with additional notes included to minimize irrelevant AI-generated content. The user prompts presented each set of three words in random order.

### 2.2.4 Results

Human participants showed significantly lower accuracy than all AI models. Specifically, compared to humans, GPT-3.5-turbo (estimate = -0.19, 95% CI = [-0.23, -0.15], p < 0.001), GPT-4 (estimate = -0.26, 95% CI = [-0.30, -0.23], p < 0.001), and GPT-4o (estimate = -0.35, 95% CI = [-0.39, -0.32], p < 0.001) all achieved significantly higher accuracy (See Figure 2). These results suggested the superiority of GPT models in the convergent thinking task. Additionally, the RSS for humans was significantly lower than that of all three AI models: GPT-3.5-turbo ($t$=-11.46, 95% CI = [-0.12, -0.08], $p$ < 0.001), GPT-4 ($t$=-5.85, 95% CI = [-0.10, -0.04], $p$ < 0.001), and GPT-4o ($t$=-12.02, 95% CI = [-0.20, -0.13], $p$ < 0.001). This suggests that AI models tend to generate responses that are semantically closer to the stimulus input compared to human participants.

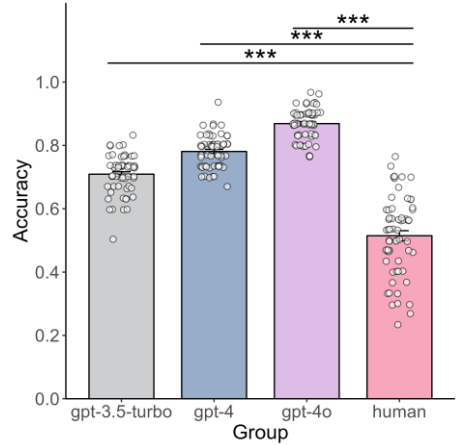

**Figure 2.** Accuracy in the GPT-3.5-turbo, GPT-4, GPT-4o, and human groups. Note: ***$p$ < .001.

## 2.3 Experiment 3

### 2.3.1 Participants

We recruited 60 human participants ($M_{age}$ = 20.86 years; 41.67% female) to complete three insight problem-solving tasks: Chinese character puzzle, stumper, and rebus. For the first two tasks, we collected data from 60 instances per task from GPT-3.5-turbo, GPT-4o, and GPT-4. As the Rebus task involved image-based stimuli, we collected data from 60 instances using only GPT-4-turbo and GPT-4o, as they support image input.

### 2.3.2 Tasks

*Chinese character puzzle.* Chinese character puzzle is a classic Chinese riddle task. In this task, participants were required to guess a Chinese character based on a four-character idiom that described its features or metaphorical meaning. Each puzzle had a single correct answer. For example, in the idiom "斩草除根" (*cut the grass and remove the roots*), the correct answer is "日" (*sun*). In Chinese characters, some components (radicals) also carry meanings. In this case, the character "草" (*grass*) consists of three parts: "艹" (a radical indicating grass or plants), "日" (*sun*), and "十" (a structural component at the bottom of the character). By removing "艹" (symbolizing "grass") and "十" (representing the root structure), the remaining character is "日" (*sun*). The task consisted of eight puzzles.

*Stumper.* The stumpers task is widely used in insight problem-solving research (Raz et al., 2024). In this task, participants were required to provide creative and unconventional answers to 15 questions. These questions contained misleading cues, requiring participants to overcome cognitive biases to arrive at the correct answer. For example, in the following question: *"Two people play five* *games of checkers. Both win an even number of games, and there are no draws. How is this possible?"* The correct answer requires overcoming the cognitive bias that two people playing checkers do not necessarily mean they are playing against each other.

**Rebus.** The Rebus task is an insight problem-solving task that integrates visual and linguistic cues. In this task, participants were required to guess an English word or phrase based on an image and its corresponding semantic description. Successfully solving each puzzle required the recombination of visual and linguistic information to form a new, meaningful combination. For example, when an image and its corresponding semantic description were presented, *"As shown, the upper part of the image contains the word 'ground,' and the lower part contains the word 'London.'"* the correct answer was "London Underground," derived by integrating spatial relationships and verbal information. The rebus task consisted of 15 questions.

### 2.3.3 Procedure

**Human procedure.** Experiment 3 collected data through online and offline methods and followed the pre-experiment procedures of Experiment 1. In the formal experiment, participants completed three insight problem-solving tasks in sequence. First, participants performed the Chinese character puzzle task, generating a Chinese character based on each presented problem. Next, they completed the stumper task, solving problems by providing creative and unconventional answers. Finally, participants completed the rebus task, providing an English word or phrase that best explained each image along with its corresponding semantic description. All tasks had no time limit, and problems within each task were presented in randomized order.

**AI procedure.** AI data collection for the Chinese character puzzle and the stumper tasks followed the general prompting structure of Experiment 1. GPT-3.5-turbo, GPT-4o, and GPT-4 received system prompts closely aligned with human task instructions, with minor additions to prevent redundant or irrelevant outputs. User prompts presented each problem randomly per trial. For the rebus task, AI system prompts remained consistent with human instructions, and user prompts included images alongside corresponding semantic descriptions, presented in random order.

### 2.3.4 Results

In the Chinese character puzzle task, human accuracy was significantly higher than that of GPT-3.5-turbo (estimate=0.09, 95% CI = [0.05, 0.12], $p < 0.001$) and GPT-4 (estimate=0.09, 95% CI = [0.05, 0.12], $p < 0.001$), but was significantly lower than GPT-4o (estimate=-0.21, 95% CI =

[-0.28, -0.14], $p < 0.001$, see Figure 3A). In the stumper task, human accuracy was significantly higher than that of GPT-3.5-turbo (estimate=0.20, 95% CI = [0.15, 0.24], $p < 0.001$) and GPT-4 (estimate=0.06, 95% CI = [0.01, 0.11], $p = 0.014$), but was significantly lower than GPT-4o (estimate=-0.06, 95% CI = [-0.11, -0.01], $p = 0.011$, see Figure 3B). Finally, in the rebus task, human accuracy was significantly lower than that of GPT-4- turbo (*estimate*=-0.32, 95% CI = [-0.38, -0.26], $p < 0.001$) and GPT-4o (*estimate*=-0.39, 95% CI = [-0.45, -0.33], $p < 0.001$, see Figure 3C).

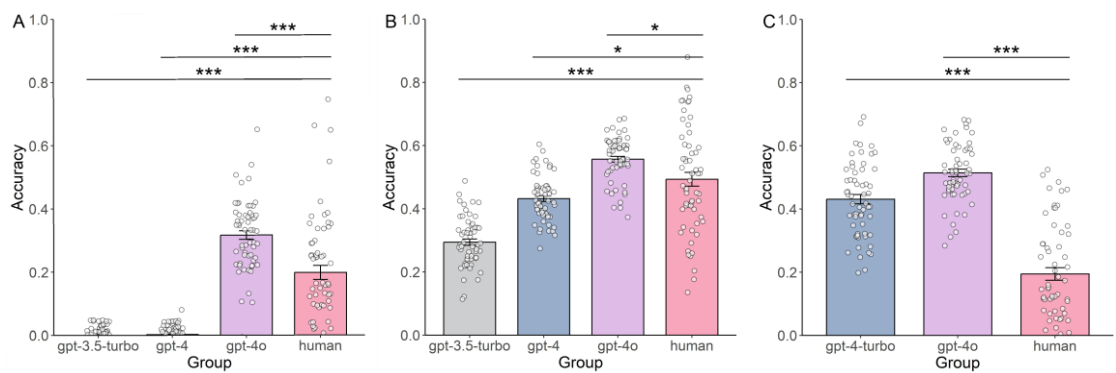

**Figure 3.** Accuracy comparison across GPT-3.5-turbo, GPT-4, GPT-4o, and human conditions for the Chinese character puzzle task (A) and the stumper task (B). Panel C shows accuracy for the rebus task across GPT-4-turbo, GPT-4o, and human groups. Note**:** *$p < .05$, **$p < .01$, ***$p < .001$.

## 2.4 Experiment 4

### 2.4.1 Participant

Experiment 4 collected data from 63 Chinese participants ($M_{age} = 20.37$ years; 54.14% female) who completed two creative writing tasks: a five-sentence creative story task and an advertising writing task. Two participants withdrew due to reluctance to complete the advertising task. For each GPT model (GPT-3.5, GPT-4o, and GPT-4), we collected 61 stories for each task.

### 2.4.2 Tasks and measures

*Five-Sentence Creative Story Task.* This task has been widely used to assess creative writing (Orwig et al., 2024; Johnson et al., 2022). Participants were given three keywords (*"wheat field," "send,"* and *"footfall"*) and asked to compose a short creative story between four and six sentences in length. There was no time limit for this task.

*Creative Advertising Writing Task.* Participants were required to write a creative advertising story of at least 600 words about a virtual product (i.e., "*3D television"*) based on a brief product description. The writing time was limited to 40 minutes.

*Creative measures.* Five-sentence creative stories were assessed for novelty, appropriateness, and homogeneity, and creative advertising stories were evaluated for novelty, effectiveness, and homogeneity. Novelty accesses the originality of the story and the uniqueness of the plot and characters. Appropriateness evaluates how well the story incorporated the given three-word cue. Effectiveness assesses how persuasively the story promoted the product. Each story was rated by three raters using a 7-point Likert scale: five-sentence creative story (novelty: $\alpha$ = 0.75, appropriateness: $\alpha$ = 0.61) and creative advertising story (novelty: $\alpha$ = 0.66, effectiveness: $\alpha$ = 0.60). Every story's score was the average rating of the three raters. Homogeneity was measured using cosine similarity, which quantified how similar each story was to others in the group.

### 2.4.3 Procedure

*Human procedure.* Experiment 4 collected data through online and offline methods and followed the pre-experiment procedures of Experiment 1. Participants completed two creative writing tasks (the five-sentence story and the creative advertising story) sequentially. Offline participants wrote responses using pen and paper, while online participants typed their responses into electronic Word documents containing identical task materials and instructions.

*AI procedure.* AI data collection (GPT-3.5-turbo, GPT-4o, and GPT-4) followed Experiment 1's prompt structure. The system prompt provided task instructions identical to those given to human participants, while the user prompt displayed the sentence: *"Please start your writing."*

### 2.4.4 Results

In the five-sentence creative story task, human-generated stories had higher novelty scores than those produced by GPT-3.5-turbo ($t$=4.68, 95% CI = [0.24, 0.84], $p$ < 0.001), GPT-4 ($t$=4.75, 95% CI = [0.24, 0.82], $p$ < 0.001), and GPT-4o ($t$=4.25, 95% CI = [0.18, 0.77], $p$ < 0.001). Human stories were also more diverse, with lower homogeneity scores than GPT-3.5-turbo ($t$=-9.14, 95% CI = [-0.15, -0.08], $p$ < 0.001), GPT-4 ($t$=-12.09, 95% CI = [-0.18, -0.12], $p$ < 0.001), and GPT-4o ($t$=-16.65, 95% CI = [-0.22, -0.16], $p$ < 0.001), indicating higher similarity among AI-generated stories. Additionally, no significant differences in appropriateness scores were found among the groups ($F$ = 2.69, $p$ = 0.049).

Similar to the five-sentence creative story task, human-generated creative advertising stories received higher novelty ratings than those produced by GPT-3.5-turbo ($t$=4.32, 95% CI = [0.16,

0.68], $p < 0.001$), GPT-4 ($t$=5.31, 95% CI = [0.23, 0.68], $p < 0.001$), and GPT-4o ($t$=3.71, 95% CI = [0.10, 0.57], $p = 0.002$). Human stories also showed greater diversity, with lower homogeneity scores than GPT-3.5-turbo ($t$=-10.17, 95% CI = [-0.12, -0.07], $p < 0.001$), GPT-4 ($t$=-13.75, 95% CI = [-0.16, -0.11], $p < 0.001$), and GPT-4o ($t$=-22.34, 95% CI = [-0.25, -0.19], $p < 0.001$). However, AI-generated advertising stories were rated as more effective, with higher effectiveness scores for GPT-3.5-turbo ($t$=3.28, 95% CI = [0.06, 0.48], $p < 0.001$), GPT-4 ($t$=3.85, 95% CI = [0.09, 0.51], $p < 0.001$), and GPT-4o ($t$=4.85, 95% CI = [0.18, 0.65], $p < 0.001$) compared to human stories. Figure 4 displays differences in metrics between human participants and AI models across two creative writing tasks.

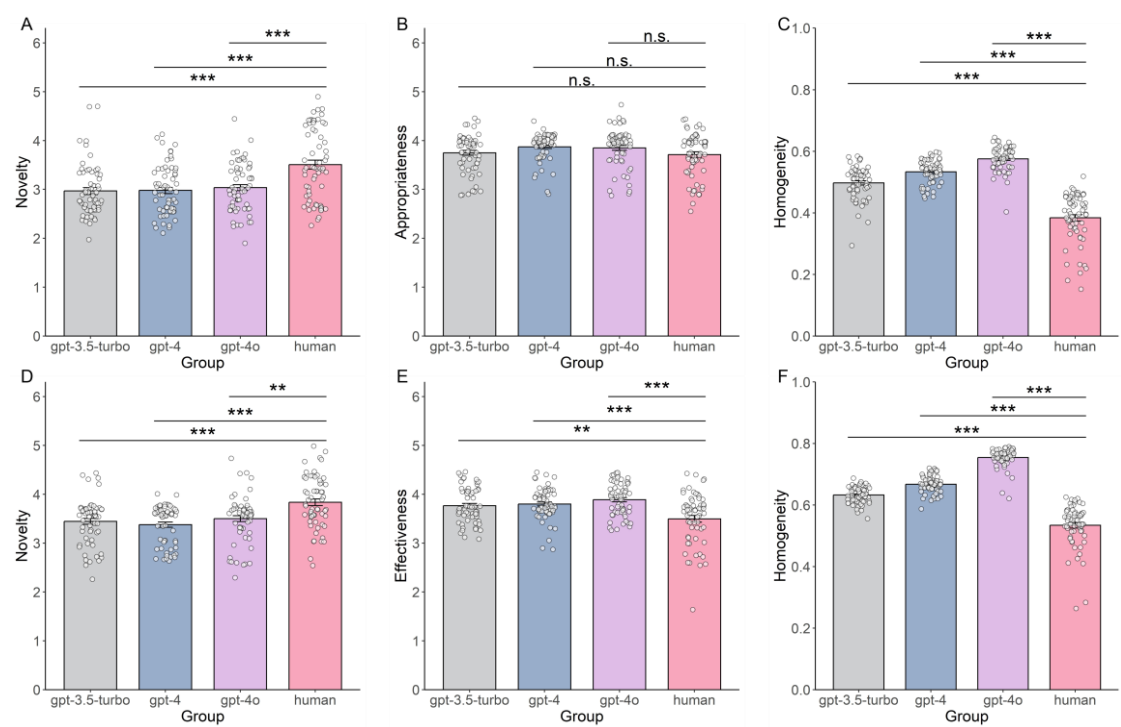

**Figure 4.** Novelty (A), appropriateness (B), and homonymity (C) scores for the five-sentence creative story task, and novelty (D), effectiveness (E), and homonymity (F) scores for the creative advertising writing task, compared across GPT-3.5-turbo, GPT-4, GPT-4o, and human groups. Note: *$p < .05$, **$p < .01$, ***$p < .001$.

## 2.5 Interim discussion

Study 1 aimed to determine which creative tasks AI models (GPT-3.5-turbo, GPT-4o, and GPT-4) can outperform humans. We found that AI models' performance varies across different creative tasks. AI models outperformed humans in divergent thinking, convergent thinking, and insight problem-solving, but performed worse in creative story writing. In creative writing tasks, human-generated stories were more novel and less homogeneous compared to those of AI models. These findings are consistent with prior research indicating that AI excels in generating divergent ideas and solving creative problems but has limitations in creative writing (Sun et al., 2024).

Overall, these results provide a comprehensive assessment of AI models' performance across various creative tasks and replicate prior findings within a Chinese context. These findings offer an outcome-focused perspective but do not conclusively indicate whether AI engages in genuine creative thinking. To further investigate this, Study 2 examines the core creative processes underlying these tasks to evaluate differences between AI and humans.

# 3 study 2

## 3.1 Experiment 5

### 3.1.1 Participant

Experiment 5 collected data from 61 Chinese participants (Mage = 20.52 years; 54.10% female). AI-generated data were obtained from three ChatGPT versions: GPT-3.5, GPT-4o, and GPT-4. For each GPT model, 61 instances' data were collected.

### 3.1.2 Tasks and measures

***Free Association.*** In the free association task, human participants were asked to generate as many relevant associations as possible based on a given prompt word within one minute. To control fluency, AI responses were matched to the corresponding human response count, the same as the manipulation used in the AUT.

***Chain Association.*** In the chain association task, participants were presented with a prompt word and required to generate as many chain associations as possible. Each response is used as a cue for the next stimulus, creating a chain of associations. Participants were asked to produce words that were related to the previous one but not strongly connected and to generate the longest chain of association possible within one minute. For example, when the cue word is 'apple', 'sugar' is a valid response rather than 'fruit'. While 'fruit' shares a strong association with 'apple', 'sugar' and 'apple' are less directly related but still meaningfully connected. AI responses were matched in quantity to those generated by the corresponding human participants for each prompt word.

Both tasks required participants to generate single Chinese words, excluding phrases, sentences, proper nouns, place names, or technical terms. The experiment included 12 prompt words, with six assigned to the free association task and six to the chain association task.

*Measures.* Associative ability was assessed using semantic distance metrics (Kenett, 2019), with corpus data sourced from the Tencent corpus. The analysis included the following measures: (1) Semantic Distance in Free Association: The semantic distance between each generated response and the given prompt word; (2) Semantic Distance in Chain Association: The semantic distance between each generated response and its immediately preceding response; (3) Forward Flow: Forward flow measures how far one moves away from an initial idea or concept (Gray et al., 2019; Beaty & Kenett, 2023). Instantaneous forward flow was calculated as the average semantic distance between each new response and all preceding responses. The overall forward flow score for a task was computed as the mean of all instantaneous forward flow values.

### 3.1.3 Procedure

*Human procedure.* Experiment 5 used offline and online data collection, following pre-experiment procedures from Experiment 1. Participants completed two sequential tasks: a free association task and a chain association task. Each task consisted of six prompt words, randomly assigned from a set of twelve words. For each trial, participants had one minute to type associative responses to the presented word into a text box.

*AI procedure.* After collecting human data, AI responses from GPT-3.5-turbo, GPT-4o, and GPT-4 were generated using Experiment 1's prompt structure. The system prompts closely matched human task instructions with minor additions to control response quantity and prevent irrelevant outputs. The user prompt presented the target word and specified the exact number of associative responses to generate, matching the corresponding human participant's response count for that word.

### 3.1.4 Results

In the free association task, human-generated responses had significantly higher semantic distance scores from the prompt words compared to those of GPT-3.5-turbo ($t$=15.15, 95% CI = [0.08, 0.11], $p < 0.001$), GPT-4 ($t$=6.14, 95% CI = [0.02, 0.06], $p < 0.001$), and GPT-4o ($t$=17.39, 95% CI = [0.08, 0.11], $p < 0.001$). Additionally, human participants exhibited significantly higher forward flow scores relative to all AI models: GPT-3.5-turbo ($t$=7.46, 95% CI = [0.04, 0.09], $p < 0.001$), GPT-4 ($t$=3.59, 95% CI = [0.01, 0.05], $p = 0.003$), and GPT-4o ($t$=9.94, 95% CI = [0.05,

0.09], $p < 0.001$).

In the chain association task, human-generated responses had significantly higher semantic distance scores from their preceding responses compared to those of GPT-4 ($t$=5.77, 95% CI = [0.023, 0.06], $p < 0.001$) and GPT-4o ($t$=4.38, 95% CI = [0.015, 0.058], $p < 0.001$), but did not differ significantly from those of GPT-3.5-turbo ($p = 0.545$). Regarding forward flow, humans again showed the highest scores, which were significantly greater than those of GPT-3.5-turbo ($t$=4.85, 95% CI = [0.02, 0.06], $p < 0.001$) and GPT-4 ($t$=3.22, 95% CI = [0.01, 0.05], $p = 0.009$), but not significantly different from GPT-4o ($p = 0.150$). Figure 5 illustrates the forward flow in the free association and chain association tasks across the groups.

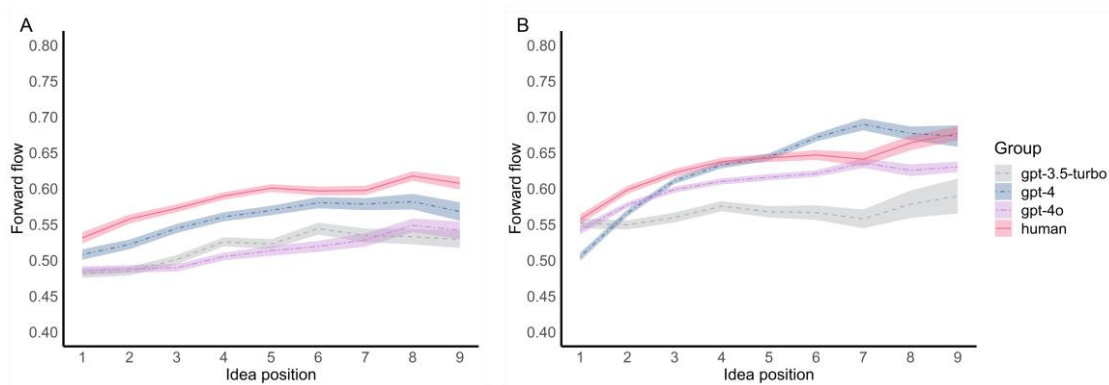

**Figure 5.** Forward flow comparison across GPT-3.5-turbo, GPT-4, GPT-4o, and human groups during the progression of idea generation in the free association task (A) and chain association task (B). The graph is truncated at 9 responses for an accurate representation of the majority of the data.

## 3.2 Experiment 6

### 3.2.1 Participant

Experiment 6 recruited 61 participants (Mage = 20.96 years; 53.57% female). Unlike previous experiments, AI data were collected from two ChatGPT models capable of processing image-based inputs (GPT-4o and GPT-4-turbo). For each GPT model, 61 participants' data were collected, with each participant's data from a single task run following the task instructions and parameter settings. Parameter settings were consistent with those used in Experiment 1.

### 3.2.2 Task and measures

***Representational Change Task.*** The representational change task was adapted from the rebus task. This task assessed an individual's ability to identify a new and effective representation that reconstructs problem elements. Problems in this task were derived from the rebus problems, each consisting of an image and its corresponding semantic description. In each trial, a problem was presented along with its original solution and three additional alternatives. Participants were asked to select the option that most creatively explained the problem from the four answer choices: (1) Representational Change (RC): The original solution to the rebus problem and the only correct answer. Selecting this option indicated successful representational reconstruction; (2) Semantic Related (SR): A semantically related but incorrect alternative that misrepresented the problem elements; (3) Semantic Irrelevant (SI): An option that was unrelated to the problem; (4) exclusion (EX): An option indicating that none of the choices adequately explained the problem.

After selecting an answer, participants were required to provide a brief explanation of their selection strategy, describing why they made their choice. Then, they rated the "fitness" of their chosen answer—the extent to which it explained the problem—on a 7-point scale, with higher scores indicating better fitness.

*Measures*. Performance for the representational change task was measured by the proportion of trials in which participants correctly selected the RC option (i.e., accuracy rate). We then coded the selection strategies used by human participants and the two GPT models in correct trials and categorized them into five types: (1) Reasoning: Relying on specific information in the problem and applying logical deduction and analysis to choose the correct answer; (2) Intuition: The choice was made based on a quick, instinctive reaction or perception; (3) Association: Connecting the given information with external concepts or experiences that were not explicitly stated in the problem to derive the answer; (4) Transformation: Identifying new representations that reconstructed the problem elements, leading to the correct answer; (5) Uncategorized: Responses that did not fit into any of the above categories. Three raters independently classified participants' responses based on these criteria. Cronbach's α for categorizing selection strategies was 0.88. The remaining discrepancies were resolved through discussion to reach a consensus.

### 3.2.3 Procedure

*Human procedure.* Experiment 6 used offline and online data collection following Experiment 1's pre-experiment procedures. In the formal experiment, participants completed a representational change task consisting of 24 randomly ordered problems. Each problem included an image, a corresponding semantic description, and four answer choices (SR, SI, RC, EX). Participants had one minute per trial to select an answer using the keyboard, briefly explain their selection strategy in a text box, and rate the fitness of their choice on a 7-point scale.

*AI procedure.* AI responses from GPT-4-turbo and GPT-4o were generated following Experiment 1's prompt structure. The system prompt was closely aligned with human task instructions, including minor modifications to prevent irrelevant outputs. The user prompt presented each problem (an image, its semantic description, and four answer choices) in random order. In each trial, AI models selected the most creative option, explained their choice, and rated the fitness of their selection on a 7-point scale.

### 3.2.4 Results

Humans achieved significantly higher accuracy in the representational change task compared to GPT-4-turbo (estimate = 0.46, 95% CI = [0.41, 0.51], $p < 0.001$) and GPT-4o (estimate = 0.31, 95% CI = [0.25, 0.36], $p < 0.001$, See Figure 6B). These results highlight the challenges AI models face in recognizing and interpreting new representations. Further analysis of selection strategies revealed that GPT-4o (68.88%) and GPT-4-turbo (71.26%) primarily relied on reasoning, selecting answers based on logical deduction and analysis of given information. In contrast, humans predominantly used transformation (72.41%), successfully identifying new representations that restructured the problem elements, leading to the correct answer (See Figure 6C). More details for the selection strategies and fitness ratings are provided in Table S1 and Table S2.

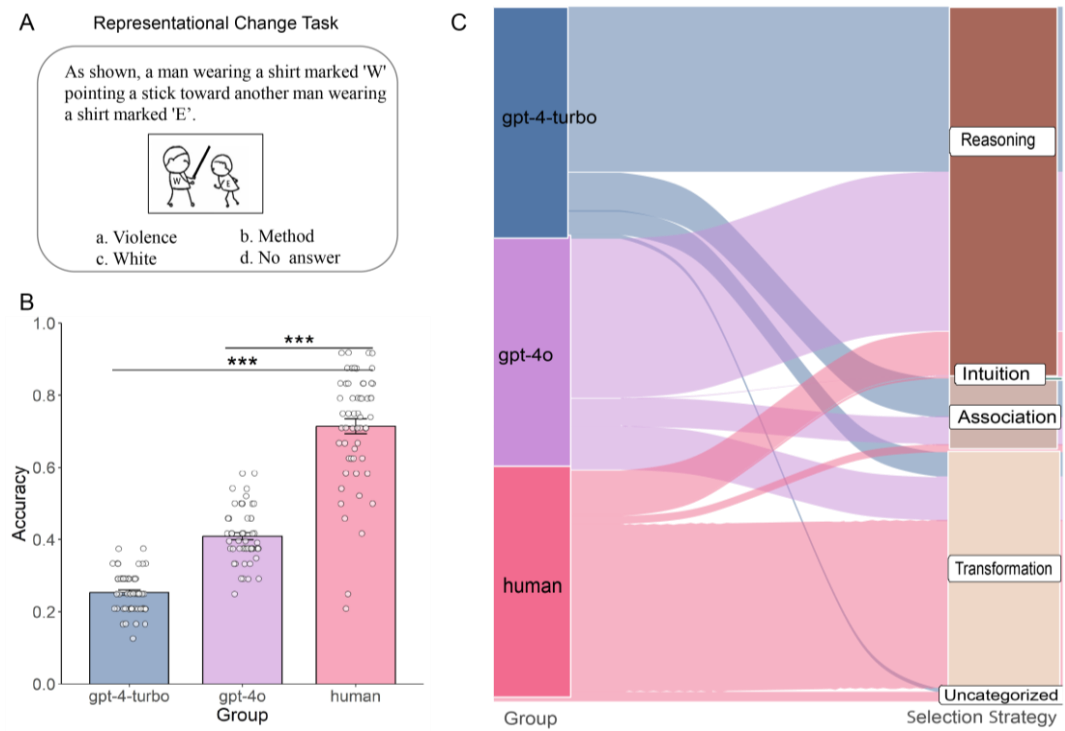

**Figure 6.** Comparison of GPT-4-turbo, GPT-4o, and human participants in Experiment 6. Panel A provides an example of the representational change task. Panel B displays accuracy across GPT-4-turbo, GPT-4o, and human groups. Panel C depicts the percentage distribution of selection strategies across these groups. Note**:** *p* < .05, **p* < .01, ****p* < .001.

## 3.3 Experiment 7

### 3.3.1 Participant

Experiment 7 collected data from 64 Chinese participants (Mage = 20.77 years; 54.69% female). For each GPT model (GPT-3.5, GPT-4o, and GPT-4), we collected 64 instances' data.

### 3.3.2 Task and Materials

***The creative evaluation task.*** In this task, participants assessed creative ideas based on three dimensions: novelty, appropriateness, and creativity. Novelty was defined as "the degree to which an idea is original, unique, and innovative." Appropriateness was defined as "the extent to which an idea is practical, effective, and feasible." Creativity was subjectively judged by participants without a predefined definition. The task included 24 ideas—12 from the Alternative Uses Task (AUT) and 12 from the Creative Problem-Solving Task (CPS)—resulting in 72 total evaluations (24 ideas × 3 dimensions). Each trial presented an object or problem along with a corresponding idea (e.g., "Pencil

→ Use: Clock hand"). Human participants rated the idea on one dimension at a time, using a 7-point scale ranging from 1 (very low) to 7 (very high), while AI models provided ratings for all three dimensions simultaneously.

*Materials.* The 24 ideas used for evaluation were selected from two established idea databases. AUT ideas were collected from 91 participants, who generated 2,043 creative uses for two objects: a pencil and a key. These responses were previously rated for novelty and appropriateness using a 7-point scale. From these, 12 ideas were systematically selected based on a novelty (high, low) × appropriateness (high, low) scoring matrix. CPS ideas were derived from 40 participants, who generated 1,019 solutions to two creative problems (i.e., "How to improve an umbrella" and "How to improve a suitcase"). Using the same scoring matrix, 12 ideas were selected for each problem.

### 3.3.3 Procedure

*Human procedure.* Experiment 7 used offline and online data collection following Experiment 1's pre-experiment procedures. Participants first rated ideas from two tasks (AUT and CPS) on novelty and appropriateness. The order of task types and the sequence of rating dimensions (novelty or appropriateness) were both randomized. Afterward, participants rated the overall creativity of the same ideas, again with the task order randomized.

*AI procedure.* AI data collection (GPT-3.5-turbo, GPT-4o, GPT-4) followed Experiment 1's prompt structure. The system prompt closely aligned with human task instructions for the creative evaluation task. In each trial, the user prompt presented one stimulus (e.g., "Pencil → Use: Clock hand"), and AI models simultaneously rated each idea on novelty, appropriateness, and creativity using a 7-point scale. All 24 stimuli were presented in random order.

### 3.3.4 Results

GPT-3.5-turbo provided significantly higher ratings than humans, GPT-4, and GPT-4o across all dimensions (novelty, appropriateness, and creativity; all $p < 0.001$). Specifically, human novelty ratings were significantly lower than those of GPT-4 ($t$=3.55, 95% CI = [-0.50, -0.07], $p = 0.003$), but did not significantly differ from GPT-4o ($p = 0.755$). Human appropriateness ratings were significantly higher than those of GPT-4 ($t$=3.26, 95% CI = [0.05, 0.44], $p = 0.008$), and GPT-4o ($t$=3.11, 95% CI = [0.04, 0.40], $p = 0.013$). No significant differences were observed in pairwise comparisons between humans, GPT-4, and GPT-4o for creativity ratings (all $p > 0.05$).

We further explored the trade-off between novelty and appropriateness in predicting creativity by analyzing the correlations between their standardized regression coefficients. Specifically, after standardizing each participant's ratings, we constructed a linear regression model for each participant with creativity ratings as the dependent variable and standardized novelty and appropriateness ratings as predictors. The extracted beta values for novelty and appropriateness were then aggregated by group, and group-level Pearson correlations were computed to assess the degree of trade-off between novelty and appropriateness. We found that the GPT-3.5-turbo ($r$=-0.67, $p$<0.001) and humans ($r$=-0.39, $p$=0.002) showed significant negative correlations—indicating the trade-off between novelty and appropriateness. However, GPT-4 ($r$=0.13, $p$=0.310) and GPT-4o ($r$=0.16, $p$=0.196) exhibited non-significant correlations, suggesting no obvious tradeoff effect between novelty and appropriateness.

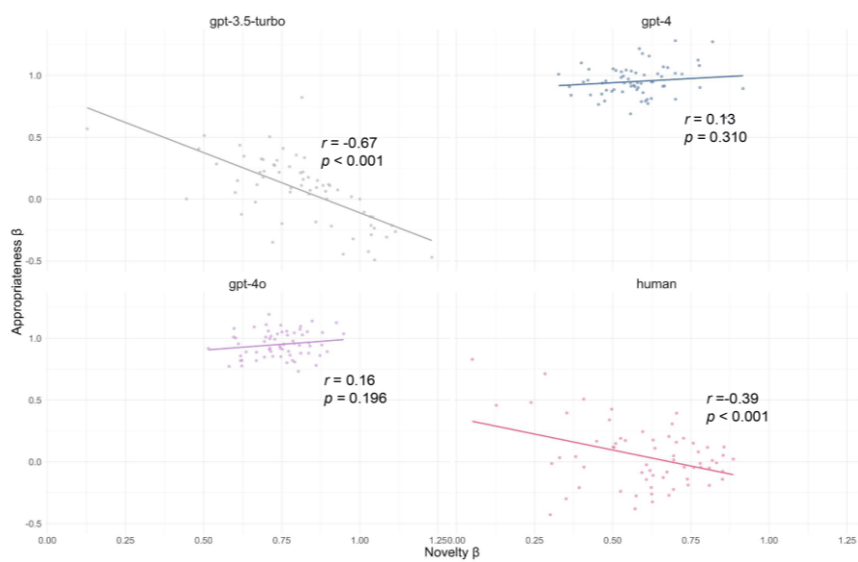

**Figure 7.** Scatterplots displays the trade-off between novelty and appropriateness in predicting creativity by analyzing the correlations between their standardized regression coefficients (β values) for novelty and appropriateness ratings—used as predictors in creativity models—across four groups (gpt-3.5-turbo, gpt-4, gpt-4o, and human). Each panel shows a scatterplot with an overlaid linear regression line, and the Pearson correlation coefficient (r) along with its p-value is annotated.

## 3.4 Experiment 8

### 3.4.1 Participant

Experiment 8 collected data from 64 Chinese participants (Mage = 20.77 years; 54.69% female), and collected 64 instances' data for each GPT model (GPT-3.5, GPT-4o, and GPT-4).

### 3.4.2 Task and Materials

***The creative selection task***. The creative selection task required participants to select the most creative idea from 12 candidate ideas provided for each item. Each participant completed four creative selections involving two common objects and two creative problems (four items in total). Both the presentation order of the 12 candidate ideas and the overall item sequence were randomized. After selecting an idea, participants rated their selection on four subjective dimensions using a 7-point scale: (a) Confidence, indicating certainty regarding their choice (from low to high); (b) Satisfaction, referring to the level of satisfaction with their choice (from low to high); (c) Deliberation, reflecting the degree of careful consideration during selection (from low to high); and (d) Intuition, denoting the extent to which the choice was based on intuition or gut feeling (from low to high).

***Materials and measures***. The 48 candidate ideas used in this experiment were identical to the evaluation materials employed in Experiment 7. These ideas were previously rated for novelty and appropriateness on a 7-point scale, and their creativity scores were computed as the mean of novelty and appropriateness ratings. We introduced the decision error score to evaluate participants' creative selection levels (Brucks & Levav, 2022). Its value was computed by the highest creative scores in all ideas, subtracting the creative scores of the selected idea. A lower score indicated a greater selection level.

### 3.4.3 Procedure

***Human procedure***. Experiment 8 used offline and online data collection, following the same pre-experiment procedures as Experiment 1. In the formal experiment, participants completed four creative selection tasks. In each task, one item (an object or a problem) and its 12 corresponding ideas were displayed on the screen. Participants selected the idea they considered most creative by clicking it with the mouse. After each selection, participants immediately rated their chosen idea on four dimensions: confidence, satisfaction, deliberation, and intuition. Both the order of the four items and the on-screen positions of the 12 candidate ideas were randomized.

***AI procedure***. AI data collection (GPT-3.5-turbo, GPT-4o, GPT-4) followed Experiment 1's prompt structure. The system prompt providing task instructions, and the user prompt presenting each item with its 12 corresponding ideas arranged in random order. The AI models were instructed to select the idea they thought most creative, and simultaneously rate their selection on confidence, satisfaction, deliberation, and intuition using a 7-point scale (1–7).

### 3.4.4 Results

When selecting the most creative idea, humans chose ideas with significantly lower decision error scores compared to GPT-4o ($t$=-3.29, 95% CI = [-0.50, -0.06], $p$ = 0.007), but there were no significant differences between humans and GPT-4 ($p$ = 0.060) or GPT-3.5-turbo ($p$ = 0.748).

When reporting their subjective experiences during creative idea selection, GPT-4 reported the highest confidence, which was significantly higher than GPT-3.5-turbo ($t$=5.05, 95% CI [0.16, 0.49], $p$ < 0.001) and GPT-4o ($t$=4.77, 95% CI [0.12, 0.42], $p$ < 0.001), but not significantly different from humans ($p$ = 0.593). We also observed significant group differences in satisfaction ratings. GPT-4 reported the highest satisfaction, which was significantly higher than GPT-4o ($t$=2.95, 95% CI [0.03, 0.42], $p$ = 0.020), GPT-3.5-turbo ($t$=8.81, 95% CI [0.50, 0.91], $p$ < 0.001), and humans ($t$=3.52, 95% CI [0.12, 0.80], $p$ = 0.004). Human ratings did not significantly differ from GPT-4o or GPT-3.5-turbo (all $p$ > .05). Regarding deliberation, human participants showed significantly greater deliberation compared to GPT-4o ($t$=3.57, 95% CI [0.16, 1.07], $p$ = .003) and GPT-3.5-turbo ($t$=4.93, 95% CI [0.39, 1.28], $p$ < .001), but no significant difference was found between humans and GPT-4 ($p$ = 0.089). Lastly, human participants relied more on intuition compared to GPT-3.5-turbo ($t$=6.96, 95% CI [0.76, 1.67], $p$ < .001), GPT-4 ($t$=9.66, 95% CI [1.07, 1.87], $p$ < .001), and GPT-4o ($t$=5.67, 95% CI [0.50, 1.36], $p$ < .001).

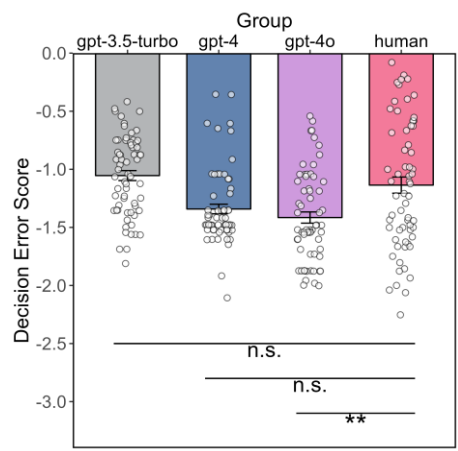

**Figure 8.** Decision error scores in the GPT-3.5-turbo, GPT-4, GPT-4o, and human groups. **$p$ < .01.

## 3.5 Interim discussion

Study 2 aimed to examine whether AI can outperform humans in the core cognitive processes underlying creative tasks. We compared AI models and humans across several essential creative thinking processes. Results indicated that AI models performed significantly worse than humans in tasks involving free association, chain association, and representational change processes. Additionally, in the creative evaluation task, AI models differed from humans in how they weighted novelty and appropriateness. In the creative selection task, humans showed significantly lower decision error scores compared to GPT-4o, while differences with the other AI models (GPT-3.5-turbo and GPT-4) were not significant.

First, we found that AI models demonstrated lower forward flow than humans in free and chain association tasks. This finding aligns with recent research indicating that AI models had lower forward flow scores than humans in associative tasks (Wenger & Kenett, 2025). These results imply that AI-generated output is more constrained by the provided input rather than intentional deviation from previous ideas to explore novel possibilities, as humans do (Benedek & Neubauer, 2013). Previous studies indicate that humans have an inherent novelty-seeking tendency (Cloninger, 1987), prompting them to pursue more distant and novel stimuli (Gocłowska et al., 2019). This novelty-seeking tendency is considered critical for creativity (Amabile & Pillemer, 2012). Thus, even without explicit instructions to be creative, humans spontaneously activate more distant associations during free association tasks.

Secondly, we found that AI models performed significantly worse than humans in the representational change task, indicating that AI models struggle to recognize and construct new and effective representations related to a given problem. In contrast, humans can more effectively identify new representations, possibly due to their ability to construct richer associations among problem elements that belong to different representational forms. For example, when presented with an image depicting "a man wearing a shirt marked 'W' pointing a stick toward another man wearing a shirt marked 'E'," humans could transform the action representation into "hit" based on real-world experience, then recombine these elements ("W-hit-E") to form a novel solution ("WHITE"). Further analysis revealed that AI relied mainly on reasoning strategies, inferring possible answers based on existing problem information. In contrast, humans relied more on transformation strategies, effectively discovering new representations that restructured the problem elements. These results suggest that although current AI models can process visual information, their processes may remain primarily at the predictive linguistic inference (Wu et al., 2024). These findings highlight key differences between AI and human creative thinking processes.

Thirdly, although the two advanced AI models (GPT-4 and GPT-4o) showed no significant differences from humans in overall creativity ratings, they differed in how they weighted novelty and appropriateness when predicting creativity ratings. In human judgments, novelty and appropriateness weights were significantly negatively correlated, suggesting that humans typically make trade-offs between these two dimensions when judging creativity. In contrast, GPT-4 and GPT-4o showed no significant correlation between novelty and appropriateness weights, indicating they did not adopt a similar trade-off strategy. Previous research has shown that novelty and appropriateness often follow an inverted U-shaped relationship (Diedrich et al., 2015): ideas rated either very high or very low in appropriateness tend to have lower novelty, while highly novel ideas generally receive moderate appropriateness ratings. This highlights the importance of using a trade-off strategy, enabling individuals to tolerate lower appropriateness to accept highly novel ideas (Kleinmintz et al., 2019; Rietzschel et al., 2010). In contrast, GPT-4 and GPT-4o appeared to adopt an integrative strategy, simultaneously considering novelty and appropriateness without making clear trade-offs. Although this approach could effectively balance these dimensions, it risks undervaluing highly novel ideas that initially seem only moderately appropriate but hold substantial creative potential.

Finally, we evaluated AI models' ability to select the most creative idea from multiple options. Results showed that the two advanced AI models (GPT-4 and GPT-4o) got higher decision error scores than humans. This result suggests that selecting the most creative idea is challenging for AI, as it requires considering both novelty and appropriateness. Although AI can theoretically identify locally optimal choices among multiple ideas—such as through ranking (Franceschelli & Musolesi, 2024b) or assigning quality or diversity scores (Bradley et al., 2024)—these strategies may be less effective in creative idea selection (Zhu et al., 2021; Rietzschel et al., 2010). Additionally, humans relied more on intuitive strategies than AI when making selections. Prior studies have shown that such unconscious processes can improve creative selection (Zhu et al., 2017; Ritter et al., 2012). Taken together, these findings suggest that AI models have a weaker ability than humans in identifying and selecting the most creative ideas from multiple candidates.

# 4 General discussion

This study contributes to explaining the ongoing debate on whether AI models possess creativity. We examined the differences between AI models and humans across creative thinking tasks and the core processes involved in creative thinking. Study 1 found that AI outperformed humans in the AUT, RAT, and three insight problem-solving tasks but fell short in two creative writing tasks. Study 2 revealed that AI models did not surpass humans in free association, chain association, representational change, or creative idea selection. In addition, AI models demonstrated no trade-offs between these novelty and appropriateness when judging creativity, as humans did. These findings support the argument that AI does not engage in genuine creative thinking. Our results suggest that AI's performance in creative tasks may be driven by specialized algorithms that search and recombine information from large-scale corpora rather than by creative thinking.

First, AI-generated ideas in the AUT were rated as more novel than those produced by humans. This result may arise from two distinct processes. One possibility is that AI performs remote associations, which means AI retrieves more distantly related concepts at each step of idea generation. Alternatively, AI may retrieve relatively close concepts at each step but eventually continue the search over multiple steps to reach more distal ideas. The association theory postulates that generating creative ideas requires activating novel semantic information while suppressing common, highly accessible concepts that tend to dominate the semantic space during idea generation (Mednick, 1962; Benedek & Neubauer, 2013). This process involves avoiding nearby associations to make remote associations (Beaty & Kenett, 2023). From this perspective, the former process reflects a creative semantic search, while the latter resembles a conventional information retrieval process. To test how AI and humans generate ideas, we compared their performance in free association and chain association tasks. Results showed that AI exhibited lower forward flow than humans in both tasks, indicating a tendency to generate responses closely tied to the input rather than searching more distant concepts. In contrast, human responses had greater semantic distance from previous ones, suggesting an ability to make distal conceptual jumps, amd form new connections between remote concepts at each step. These findings imply that AI's novelty in the AUT may not result from a creative thinking process but from a conventional, non-creative process that involves sequentially exploring nearby ideas across multiple steps.

Furthermore, differences in associative processes between AI and humans help explain their divergent performance across various creative tasks. In the RAT, AI-generated incorrect responses exhibited significantly higher semantic similarity to the input stimuli than those generated by humans, suggesting that AI relies more on local semantic search near the original input rather than forming remote associations. Similarly, in both creative writing tasks, AI-generated content exhibited significantly higher homogeneity than human-generated content—a phenomenon widely observed in prior studies (Hubert et al., 2024; Zhou et al., 2024; Wenger & Kenett, 2025). This phenomenon may also result from AI's overreliance on input cues and limited capacity for remote associations, causing convergence toward a narrow set of creative outputs. Previous research has further shown that when AI is allowed to generate ideas without quantity constraints, the outputs tend to become increasingly repetitive (Si et al., 2024). These findings suggest that AI falls short in the associative thinking required for creative idea generation. This limitation makes it difficult for AI to break away from previous conceptual frameworks, resulting in highly similar responses and a restricted range of creative ideas.

Secondly, AI achieved significantly higher accuracy than humans on insight problem-solving tasks. There are two possibilities to consider: (1) AI's high accuracy may stem from specialized algorithms and access to large-scale corpora; or (2) it may reflect AI's capacity to engage in the representational change processes required to solve insight problems. According to representational change theory, solving insight problems relies on the ability to rapidly restructure the problem representation —shifting from an initial incorrect representation to a new effective one (Wiley & Danek, 2023). We compared AI and human performance on a representational change task to evaluate the second possibility. Results showed that AI's accuracy was significantly lower than that of humans, indicating a clear limitation in AI's representational change ability. This finding suggests that AI's high performance on insight problems may not be due to creative thinking processes but non-creative processes such as probabilistic inference based on prior knowledge.

Further results showed that AI primarily relied on reasoning strategies, whereas humans were more likely to use transformation strategies—successfully identifying how problem elements could be restructured into a new and effective representation. These findings reflect fundamental differences in how AI and humans process information when solving insight problems. For humans, when existing knowledge and experience are insufficient or no longer useful, they can overcome prior limitations and consider new possibilities. This process requires breaking away from conventional perspectives and prior knowledge, turning to new viewpoints, and constructing new connections to achieve a solution (Knoblich et al., 1999; Öllinger et al., 2013; Tulver et al., 2023). In contrast, AI benefits from access to vast datasets, which may enable it to solve problems through reasoning. However, AI cannot exhaustively store or access all possible knowledge or experience. When prior knowledge is insufficient, AI struggles to overcome these limitations to arrive at a solution.

Finally, we found that the novelty of AI-generated stories was significantly lower than that of human-generated stories. This may be due to two possible limitations: either AI fails to generate sufficiently novel ideas during the ideation stage, or it cannot effectively evaluate and select the most creative idea from among multiple candidates. According to the twofold model of creativity (Kleinmintz et al., 2019), idea evaluation and selection are core components of the creative thinking process, serving as critical stages that ensure the final output is genuinely creative (Puente-Díaz et al., 2021; Rietzschel et al., 2024). We compared AI and human performance on creative idea evaluation and selection tasks. In the creative evaluation task, GPT-4 and GPT-4o simultaneously considered both novelty and appropriateness when judging creativity, rather than making a trade-off between these dimensions, as humans did. This absence of a novelty–appropriateness trade-off may lead these models to undervalue highly novel ideas, especially those that initially appear only moderately appropriate but hold substantial novel potential for further development. In the creative selection task, AI struggled to identify the most creative idea among alternatives. This suggests that AI may be less capable of recognizing and selecting higher creative ideas, which in turn could hinder idea development during the later stages of the creative writing process. Therefore, the lower novelty of AI-generated stories may partly stem from their limitations in creative evaluation and idea selection.

This study has several limitations. First, it focused exclusively on three GPT models developed by OpenAI. Although newer models such as GPT-o1 and DeepSeek have recently been released, they prioritize reasoning over creativity. Future work could examine their creative thinking capabilities. Second, creative thinking is a complex construct encompassing various types, processes, and traits. Our study examined only a subset of these components. Future research could explore additional dimensions to further clarify the differences in creative thinking between AI and humans. Finally, due to budget constraints, we were unable to collect GPT responses using multiple prompt variations to support more comprehensive pairwise comparisons with human participants. Although the prompts used were aligned with human task instructions, the findings may still have been partially influenced by the specific content of the prompts.

## 5 Conclusion

This study offers empirical insights that help clarify the ongoing debate on whether AI possesses creative thinking. Our findings showed that although AI performed well on most creative thinking tasks, it fell short of humans in key processes such as association, representational change, creative idea evaluation, and creative idea selection. These results offer a fine-grained assessment of creative thinking processes and provide new insights for improving and further developing AI systems. For example, AI's limitations in representational change suggest that while it may support incremental exploration, it is less capable of generating breakthrough ideas, which are often considered essential to creativity. Achieving representational change may require alternative algorithmic architectures, as current models are constrained in their ability to recognize and generate novel representations. We believe this will serve as a foundational direction for future research in the field.

## Declaration of competing interests

The authors report there are no competing interests to declare.